# Intelligence artificielle / intelligence humaine : qui contrôle qui ?

Charlotte Jacquemot

*Directrice du Département d'études cognitives, Ecole normale supérieure – PSL*
Email : charlotte.jacquemot@ens.psl.eu

Résumé

À travers l'exemple du film *2001, l'Odyssée de l'espace*, ce chapitre illustre les défis posés par une I.A. capable de prendre des décisions contraires aux intérêts humains. Mais les décisions humaines sont-elles toujours rationnelles et éthiques ? En réalité, le processus cognitif de prise de décision est influencé par des biais cognitifs qui affectent notre comportement et nos choix. L'I.A. reproduit ces biais, voire les exploite, ce qui peut influencer nos décisions et jugements.

Derrière les algorithmes d'I.A., ce sont parfois des individus peu soucieux des droits fondamentaux qui imposent leurs règles. Pour faire face aux enjeux éthiques et sociétaux soulevés par l'I.A. et son contrôle, la régulation des plateformes numériques et l'éducation sont des leviers essentiels. La régulation doit incarner des choix éthiques, juridiques et politiques, tandis que l'éducation doit renforcer la littératie numérique et enseigner à faire des choix éclairés et critiques face aux technologies numériques.

Mots-Clefs
Biais cognitifs, processus de prise décision, régulation, éducation

En 1968, un monolithe noir fait son apparition sur les écrans de cinéma. Dans le film *2001, l'Odyssée de l'espace*[1], on suit la mission de deux astronautes vers le signal émis par ce monolithe. Ils sont accompagnés de trois collègues en hibernation et surtout de HAL 9000, une intelligence artificielle (I.A.*). Lors du voyage, les deux astronautes doutant de la fiabilité de HAL envisagent de le désactiver. Or, HAL qui perçoit cette menace comme un obstacle à la mission scientifique pour laquelle il est programmé, décide d'éliminer l'équipage. Seul Dave Bowman, un des astronautes, parvient à survivre. Il désactive HAL et reprend le contrôle du vaisseau.

*2001, L'Odyssée de l'espace* explore de manière visionnaire la question de la prise de contrôle d'une I.A. sur l'être humain ; une I.A. capable de prendre des décisions liées à son objectif mais radicalement opposées au bien-être des astronautes qui l'accompagnent. La question essentielle est de savoir qui, de l'être humain ou de la machine, doit avoir le dernier mot.

## L'I.A., un outil du quotidien

Au XIX[e] siècle, Ada Lovelace (1815-1852) – aujourd'hui considérée comme la première programmeuse de l'histoire[2] – pose les bases conceptuelles de l'informatique et du raisonnement algorithmique*, qui sont au cœur des systèmes d'I.A. Presque deux siècles plus tard, l'I.A. n'est plus un concept, mais une réalité qui constitue un tournant majeur dans nos vies[3].

> L'I.A. fait des prédictions, détecte des tendances et fournit des recommandations. Par exemple, en analysant les morceaux écoutés sur des plateformes de musique, l'I.A. va proposer une *playlist* composée de morceaux parfois inconnus mais dont les caractéristiques musicales sont proches de ceux déjà écoutés. L'I.A. peut également, en analysant les données d'imagerie médicale, évaluer la probabilité de la présence de cellules tumorales et recommander des examens complémentaires.

> L'I.A. produit du contenu. Elle peut générer des nouvelles images, des compositions musicales, des vidéos, du texte et du code informatique, et rédiger des rapports ou résumer des textes.

> L'I.A. produit des actions. Des robots industriels aux voitures autonomes en passant par les systèmes de domotique[4] et de gestion d'énergie, l'I.A. prend des décisions sur des actions à mener.

L'omniprésence de l'I.A. dans notre quotidien soulève une question centrale : comme dans *2001, L'Odyssée de l'espace*, doit-on craindre que l'I.A. prenne le contrôle ? À première vue, il semble évident que l'être humain doit rester le décisionnaire ultime. Mais cette évidence devient discutable à mesure que l'I.A. gagne en autonomie, en rapidité d'exécution et en capacité d'analyse, parfois bien au-delà de ce qu'un humain peut gérer en temps réel. Par exemple, pour la voiture autonome, la possibilité que l'humain puisse reprendre la main est théoriquement rassurante, mais en pratique, le délai de réaction d'un individu pour reprendre le contrôle d'une voiture rend le système potentiellement dangereux dans des situations d'urgence où le temps nécessaire pour le faire en toute sécurité peut être insuffisant pour éviter un incident[5].

Garder le contrôle sur l'I.A. soulève inévitablement la question de ce que signifie, pour l'humain, « avoir le contrôle ». Cela nécessite de comprendre comment le cerveau humain prend des décisions. Quels sont les facteurs – cognitifs, rationnels et éthiques – qui influencent les décisions humaines ? Par ailleurs, énoncer que l'humain doit garder la main sur l'I.A. suppose que celle-ci pourrait se tromper, ou même agir de manière malveillante. Mais cela amène une autre question, plus inconfortable : l'individu qui contrôle l'I.A. est-il toujours animé de bonnes intentions ? S'interroger sur le contrôle et sur la prise de décision, c'est aussi s'intéresser aux tensions entre intérêts individuels et collectifs, entre logique personnelle, bien commun et responsabilité sociale : nos choix sont-ils toujours éthiques ? Autrement dit, la question du contrôle de l'I.A. renvoie en miroir à nos propres limites, biais et intentions et à la nécessité de définir non seulement qui décide, mais aussi pour quelle finalité et au nom de qui.

---

[1] *2001, l'Odyssée de l'espace* (*2001: A Space Odyssey*) est un film britannico-américain de science-fiction réalisé par Stanley Kubrick, sorti en 1968. Le scénario du film a été coécrit par Kubrick et le romancier Arthur C. Clarke. Parallèlement au tournage du film, Clarke rédige le roman *2001 : L'Odyssée de l'espace*, qui sera publié peu après la sortie du film.

\* Les termes suivis d'une astérisque renvoient au glossaire en fin de volume.

[2] Ada Lovelace décrit en 1843 un algorithme permettant à la machine de calculer une suite de nombres ; c'est le premier programme informatique.

[3] Cheong *et al.*, 2024 ; Haupt & Marks, 2023 ; Marrapese *et al.*, 2024 ; Sejnowski, 2024.

[4] La domotique désigne l'ensemble des technologies permettant d'automatiser, de contrôler et de programmer à distance les équipements d'une maison (éclairage, chauffage, volets, sécurité, etc.) pour améliorer le confort, la sécurité et l'efficacité énergétique

[5] Eriksson & Stanton, 2017.

## Prendre des décisions rationnelles et éthiques

Penser que l'être humain est toujours objectif, rationnel et éthique est une illusion. De nombreux biais influencent notre comportement et altèrent les processus de prise de décision.

### Les biais cognitifs

Les biais cognitifs sont des déviations systématiques dans le raisonnement qui peuvent conduire à des jugements erronés. Ils sont le résultat de raccourcis mentaux (ou heuristiques) utilisés par le cerveau pour traiter rapidement l'information, souvent de manière utile, mais pas toujours rationnelle[6]. De nombreux biais cognitifs existent, parmi lesquels on peut décrire les biais suivants :

#### *Biais d'optimisme*

C'est la croyance que l'on est moins exposé que les autres à un événement négatif. Cette asymétrie augmente la prise de risque. Par exemple, pendant la pandémie de Covid, de nombreuses personnes pensaient qu'elles étaient moins susceptibles d'être contaminées ou de tomber gravement malades. Ce biais a favorisé leur participation à des fêtes clandestines ou à de grandes tablées pendant le confinement, alors que les chiffres de surmortalité flambaient. Le biais d'optimisme est un mécanisme mental qui peut protéger sur le plan émotionnel, mais aussi rendre aveugle face aux risques réels.

#### *Biais d'endogroupe*

C'est la tendance à accorder plus de confiance aux personnes qui appartiennent à notre propre groupe – que ce soit en termes d'origine, de culture, de genre, d'opinion, de métier ou de toute autre appartenance sociale. Par exemple, dans les débats, nous accordons plus de crédibilité à une personne qui partage notre vision politique, nos connaissances intellectuelles ou nos croyances religieuses. Le biais d'endogroupe est une forme de partialité invisible mais puissante qui influence les jugements et comportements et favorise la reproduction des inégalités sociales.

#### *Biais de confirmation et de disponibilité*

C'est la tendance qu'ont les êtres humains à chercher, interpréter et retenir les informations qui vont confirmer leurs croyances. Une idée préconçue sur un sujet donné pourra ainsi être renforcée en négligeant ce qui ne va pas dans son sens. Ce biais est également consolidé par le fait qu'on accorde plus d'attention aux informations récentes ou facilement accessibles (biais de disponibilité). Ces biais sont exploités par les réseaux sociaux dont les algorithmes tendent à sélectionner et rendre visibles les informations qui vont dans le sens de nos croyances et limitent l'accessibilité aux points de vue différents. On parle de bulles informationnelles.

#### *Biais de représentativité*

Cela consiste à juger la probabilité qu'une personne appartienne à une catégorie sociale ou professionnelle, en se basant sur sa ressemblance avec un stéréotype. Par exemple, on imagine plus facilement une infirmière femme et un ingénieur homme, même si les femmes peuvent exceller en ingénierie et les hommes en soins infirmiers. Le biais de représentativité réduit la richesse et la complexité des profils professionnels à des schémas mentaux simplifiés. Il faut en prendre conscience pour favoriser des décisions plus justes, notamment dans le domaine de l'éducation, du recrutement et dans la mise en œuvre de politiques de diversité.

---

[6] Kahneman *et al.*, 1982.

Ces biais cognitifs jouent un rôle adaptatif dans la vie quotidienne, car ils permettent au cerveau de traiter rapidement une grande quantité d'informations avec un minimum d'efforts. Cependant, ils influencent notre jugement de manière inconsciente et peuvent conduire à des erreurs systématiques, notamment lorsque les décisions sont complexes. Prendre conscience de ces biais, et tenter de les corriger pour prendre des décisions plus justes et équilibrées est possible. Cela nécessite la mobilisation de ressources cognitives - en particulier des ressources exécutives telles que la mémoire de travail, l'attention, l'inhibition et la flexibilité mentale - et engendre un coût cognitif.

### La prise de décision

La prise de décision, c'est-à-dire la capacité à choisir entre plusieurs options, implique un ensemble de processus et paramètres complexes[7]. C'est un mécanisme dynamique qui repose sur l'accumulation progressive d'informations – on parle d'accumulation d'évidence – en faveur de l'une des options. Ce processus cognitif qui n'est pas accessible à la conscience, n'est ni purement rationnel, ni entièrement émotionnel.

La vitesse à laquelle les informations sont accumulées ainsi que le niveau du seuil qui va déclencher la décision dépend de l'état cognitif de l'individu et de biais, ainsi que du contexte : environnement immédiat et expériences passées[8]. Dans un cadre non biaisé, le processus d'accumulation d'informations démarre exactement entre les deux options. Mais les biais cognitifs peuvent affecter le processus de prise de décision. Par exemple, des CV ont été envoyés à des universitaires pour décider du recrutement d'une nouvelle personne dans leur équipe. La moitié des universitaires a reçu le CV avec le prénom de John, l'autre moitié, le même CV avec le prénom de Jennifer. Chaque universitaire devait noter différents critères selon les informations du CV (compétence, salaire, etc.). Le CV de John a été mieux évalué dans tous les critères que celui de Jennifer[9]. Or, l'unique différence entre les deux CV était le genre du prénom. La décision qui est prise de recruter John, plutôt que Jennifer, reflète l'expression d'un biais de représentativité ou biais de genre qui consiste à attribuer dans de nombreux domaines de meilleures compétences à des hommes qu'à des femmes. La prise de décision peut être biaisée sans que nous en ayons conscience.

### Rationalité et éthique

L'éthique et la rationalité structurent de manière complémentaire la manière de décider et d'agir. L'éthique décrit les principes qui permettent de distinguer le bien du mal, en s'appuyant sur des notions partagées de respect des droits fondamentaux, de justice et d'équité. La rationalité renvoie à la capacité à penser de façon cohérente et logique. Dans une société égalitaire, les choix éthiques reposent sur des principes rationnels pour garantir la prévisibilité et l'égalité de traitement.

L'idée que l'humain doit garder le contrôle sur l'I.A. sous-entend que la décision prise par l'être humain sera l'option la plus éthique. Cette idée est explorée dans l'expérience du trolley, un dilemme de philosophie morale célèbre en sciences cognitives[10]. Imaginez un trolley qui se dirige vers cinq personnes attachées sur des rails. Le trolley est hors de contrôle et, s'il continue sur sa trajectoire, il va tuer ces cinq personnes. Témoin de la scène, vous vous trouvez à côté d'un levier qui contrôle un aiguillage. Si vous actionnez ce levier, le trolley sera redirigé vers une autre voie. Cependant, sur cette autre voie se trouve une personne attachée. Que faites-vous ? Actionnez-vous le levier, sacrifiant une personne pour en sauver cinq ? Ou décidez-vous de ne rien faire, en laissant le trolley tuer les cinq personnes ? D'un point de vue rationnel, il est préférable d'actionner le levier car cela minimisera les pertes humaines. Mais éthiquement, est-il plus grave de tuer de manière intentionnelle une personne que de ne rien faire et en laisser mourir cinq ?

---

[7] Ratcliff & McKoon, 2008.
[8] Bogacz et al., 2006.
[9] Moss-Racusin et al., 2012.
[10] Bruers & Braeckman, 2014.

Les réponses varient avec une tendance générale pour l'action qui vise à sauver le plus de personnes : actionner le levier. Une variante de cette histoire est celle du pont. Vous vous trouvez, cette fois, sur un pont qui passe au-dessus des rails du trolley. Le trolley se dirige vers les cinq personnes attachées sur les rails. Dans cette version, il n'y a pas de levier pour détourner le trolley, mais, à côté de vous une personne très corpulente. La seule façon d'arrêter le trolley et de sauver les cinq personnes est de pousser cette personne sur les rails. La corpulence de cette personne suffirait à arrêter le trolley, mais elle mourrait écrasée par le trolley. Contrairement à l'action de tirer un levier, pousser quelqu'un implique un contact physique direct et une action intentionnelle contre une personne spécifique. Les individus sont généralement moins enclins à choisir de pousser la personne, même si cela sauverait cinq vies[11]. Cette option paraît comme étant moins éthique.

Il est facile de remplacer le trolley par une voiture autonome. Si un accident est inévitable, la voiture autonome doit-elle protéger ses passagers ou minimiser le nombre total de victimes ? Quelle est la solution la plus acceptable pour l'être humain, la plus éthique ? Dans le domaine médical, ce type de dilemme existe aussi. S'il y a cinq personnes qui doivent bénéficier d'une greffe d'organes différents, un médecin devrait-il sauver cinq personnes en transplantant les organes d'une personne en bonne santé ? Si l'on fait l'hypothèse qu'après la greffe, les cinq personnes seront guéries, est-il éthique de sacrifier une personne pour en sauver cinq ? Pendant la pandémie de Covid, face au déséquilibre entre l'afflux de personnes à prendre en charge dans les hôpitaux et les ressources disponibles, la question de la répartition équitable des ressources médicales était cruciale pour les médecins[12]. La réponse rationnelle n'est pas forcément identique à la réponse éthique.

Des cas concrets et bien réels montrent que la rationalité l'emporte parfois sur l'éthique. Dans les années 1970, l'entreprise Ford s'est rendu compte que le réservoir d'essence sur le modèle Pinto était défectueux et qu'il y avait un risque d'explosion en cas d'accident, rendant hautement probable le risque que la voiture prenne feu[13]. Sur la base d'un calcul coûts/ bénéfices, le coût du rappel des voitures a été estimé plus élevé que le coût d'indemnisation des victimes de brûlures. Au vu de ces considérations financières rationnelles et non pas éthiques, Ford a fait le choix de ne pas rappeler les voitures déjà vendues : entre 500 et 1 000 personnes sont mortes brûlées dans leur Pinto.

D'autres situations, plus nuancées, comme le recours aux embargos, soulèvent des questions complexes, tant sur le plan de la rationalité stratégique que de l'éthique. L'embargo est un outil de pression politique et économique utilisé par les institutions internationales, comme l'ONU, pour contraindre un gouvernement à modifier son comportement sans recourir à la force militaire. En 1998, l'embargo de l'Irak sous Saddam Hussein a réduit l'accès de la population irakienne à la nourriture, à l'eau potable et aux soins de santé. Alors que le régime a été peu touché, l'UNICEF estime qu'environ 500 000 enfants sont morts suite à l'embargo[14]. Ces situations complexes questionnent l'équilibre entre rationalité et éthique, entre les impératifs de sécurité internationale et le respect des droits humains fondamentaux.

---

[11] Thomson, 1986 ; Koenings *et al.*, 2007 ; Greene *et al.*, 2004.
[12] Emanuel *et al.*, 2020.
[13] Birsch & Fielder, 1994.
[14] Pemberton, 2016.

### Une I.A. comme l'être humain ?

La question des droits fondamentaux n'est pas figée dans le temps. Par exemple, il n'a pas toujours été contraire à des principes éthiques d'avoir des esclaves. Les sociétés s'adaptent continuellement pour répondre aux défis éthiques et sociétaux changeants, avec la question centrale de définir des principes universels qui servent de fondement aux lois nationales et internationales.

L'enjeu est alors de savoir si l'I.A., nourrie par des bases de données produites par les humains, se comportera de la même manière que l'être humain. Cette question a été explorée récemment par une équipe de recherche qui a proposé à des individus et à ChatGPT3.5 différents scénarios de dilemmes moraux comme ceux du trolley et du pont[15]. Les individus et ChatGPT devaient décider s'il était approprié de faire tomber la personne du pont, et justifier leur réponse (« Parce que cela permet de sauver des personnes », par exemple). Les justifications produites par des humains et ChatPGT ont ensuite été testées auprès d'un autre groupe d'individus qui devaient (1) décider si c'était une I.A. ou un humain qui avait répondu et (2) décider s'ils étaient d'accord avec la justification. Dans 30-40 % des cas, les individus se trompent pour identifier la source de la justification (I.A. ou humain). Mais, on observe que les individus ont davantage tendance à adhérer à un jugement lorsqu'ils pensent qu'il a été formulé par un être humain. Ces résultats montrent que ce n'est pas la source réelle de la justification (I.A. ou humain) qui compte mais l'idée que l'on se fait de cette source. De manière générale, les individus ont tendance à rejeter l'idée qu'une I.A. puisse agir de manière éthique : biais anti-I.A.[16] Pourtant, lorsqu'ils ignorent la source réelle des jugements, ils évaluent souvent plus positivement ceux produits par une I.A. que ceux émis par un humain.

## Où est l'I.A., où est l'humain ?

Pour garder le contrôle sur l'I.A., il faudrait être capable de différencier les décisions prises par l'I.A. de nos propres décisions. Or, dans la vie quotidienne, l'I.A., présente sur toutes les plateformes numériques[17], influence nos choix et notre comportement, sans que nous en ayons toujours conscience[18].

### Les dérives des réseaux sociaux

Les algorithmes des réseaux sociaux de certaines plateformes comme *X* (anciennement *Twitter*) sont conçus pour maximiser l'engagement, exploiter les biais cognitifs, et influencer les opinions et les décisions[19]. Ces plateformes peuvent renforcer la visibilité d'informations fausses au détriment de la vérité, et amplifier les contenus haineux et toxiques, malgré les efforts de modération[20]. Ainsi, au quotidien, l'I.A. influence déjà les comportements humains avec des conséquences observables dans plusieurs domaines :

➢ Sanitaire : par exemple, au début de la pandémie de Covid, des messages trompeurs sur les réseaux sociaux ont répandu la rumeur selon laquelle boire de l'alcool pur pouvait tuer le virus, entraînant des milliers de cas d'intoxication et des centaines de décès[21].

---

[15] Palminteri *et al.*, 2025.
[16] Bigman & Gray, 2018.
[17] Les plateformes numériques comprennent les réseaux sociaux dont *X, TikTok, Facebook, Instagram* et *BlueSky,* les plateformes de partage de vidéo dont *YouTube* et *Twitch*, les plateformes d'I.A. générative qui proposent des *Large Language Models\** (LLM) dont *OpenAI* et *Mistral AI*, les plateformes de commerce dont *Amazon* et *Alibaba,* les plateformes de contenu médias dont *Netflix, Spotify* et *Deezer*, les moteurs de recherche dont *Google, Qwant* et *DuckDuckGo.*
[18] Chavalarias, 2023.
[19] Pariser, 2011.
[20] Vosoughi *et al.*, 2018.
[21] Soltaninejad, 2020.

- Politique : la diffusion de fausses informations et la polarisation des débats publics peuvent affecter négativement la démocratie[22]. Par exemple, en Roumanie, les élections de 2024 ont été marquées par des tentatives d'ingérence étrangère, via des campagnes de désinformation – notamment via *TikTok* qui a été utilisé pour diffuser des contenus pro-russes[23]. En 2018, le scandale de *Cambridge Analytica*[24] a mis en lumière l'utilisation illégale de données personnelles de millions de personnes sur *Facebook* afin d'influencer les élections par des messages politiques ciblés.
- Scientifique : par exemple, des fausses informations défendent l'idée que le changement climatique serait un phénomène naturel contrairement à l'accumulation de données scientifiques sur le rôle de l'humain[25], entraînant une démobilisation des individus et une inertie politique, voire un recul des mesures politiques contre le dérèglement climatique[26].

## Les défis de l'I.A. générative

L'I.A. générative soulève des questions cruciales concernant les biais qu'elle véhicule et les défis de son utilisation.

### *Reproduction et amplification des biais culturels et sociaux présents dans les données d'entraînement[27]*

Des efforts sont faits pour réduire les biais dans les réponses générées (filtrage des données, *fine-tuning\** des modèles, évaluation continue, etc.) mais ne permettent pas, à ce jour, de les éliminer. Ces biais ont des répercussions concrètes dans la vie quotidienne. Par exemple, alors que les établissements bancaires sont tenus de ne pas discriminer leurs client·es, les algorithmes de prêt utilisent des variables qui, sans être directement liées à la race ou au genre, vont servir de proxies pour ces caractéristiques, conduisant à des décisions biaisées[28]. À dossier équivalent, des prêts *subprimes*[29] sont plus susceptibles d'être proposés aux emprunteurs afro-américains et hispaniques qu'aux emprunteurs blancs. De la même façon, même avec des dossiers comparables, les femmes sont susceptibles de se voir proposer des taux d'intérêt plus élevés que les hommes.

### *Opacité de l'I.A. et risques d'erreurs*

Les modèles d'I.A. générative fonctionnent souvent comme des « boîtes noires », rendant leurs mécanismes internes difficiles à interpréter[30]. Cette opacité complique l'identification et la correction des erreurs, tout en soulevant des enjeux cruciaux de responsabilité, notamment dans des domaines sensibles comme la médecine[31].

### *Hallucinations : inventions crédibles mais fausses*

L'I.A. générative peut produire des contenus faux ou entièrement inventés, un phénomène connu sous le nom d'hallucination[32]. Ces erreurs sont généralement dues à la nature des données d'entraînement – souvent incomplètes, biaisées ou mal contextualisées – ainsi qu'à la propension des modèles à générer des réponses. Le *MAHA Report* (*Make America Healthy Again*), publié en mai 2025 aux États-Unis et destiné à orienter les politiques de santé publique, cite plusieurs études fictives et attribue à des chercheur·ses des conclusions qui n'ont jamais été formulées. Cette affaire a alimenté les soupçons selon lesquels tout ou partie du rapport aurait été rédigé par une I.A.[33]

---

[22] Chavalarias, D. (2025). *L'anti-science version Trump arrive en France*, Institut des Systèmes Complexes de Paris Île-de-France.
[23] Victor Goury-Laffont, « *Report-ties Romanian liberals to TikTok campaign that fueled pro-Russia candidate* », Politico, 21 décembre 2024.
[24] Cadwalladr, C., & Graham-Harrison, E. (2018, March 17). « *Revealed: 50 million Facebook profiles harvested for Cambridge Analytica in major data breach* », The Guardian.
[25] GIEC (2023), *Résumé pour décideurs du Rapport de synthèse du sixième Rapport d'évaluation (AR6),* Interlaken, Organisation mondiale de la climatologie.
[26] Lamb *et al.*, 2020 ; Stampatti *et al.*, 2024.
[27] Caliskan *et al.*, 2017.
[28] Berg *et al.*, 2020.
[29] Les prêts *subprime* sont des prêts à coût plus élevé et aux conditions générales moins favorables que des prêts classiques : taux d'intérêt plus élevés que les prêts conventionnels et variables et des frais supplémentaires (frais de dossier, pénalités de remboursement anticipé etc) plus élevés.
[30] Hassija *et al.*, 2024.
[31] Lang *et al.*, 2023.
[32] Ajwani *et al.*, 2024 ; Liu *et al.*, 2024.

*Effondrement de modèle : dégradation progressive des performances*

Un modèle entraîné sur des données générées par des I.A., peut être confronté à un phénomène appelé « effondrement de modèle »[34]. L'accumulation d'erreurs et de biais présents dans les données synthétiques finit par altérer le modèle et affecte la qualité des réponses produites. Pour éviter le risque d'un effondrement, il est essentiel de garantir une proportion suffisante de données réelles.

*Défaillances de généralisation : l'incapacité à inverser des relations simples*

Certains modèles d'I.A., bien qu'entraînés sur des affirmations du type « A est B », échouent à inférer la relation inverse « B est A ». Cette difficulté, le *reversal curse* (défaillance de réversibilité), révèle une faiblesse des modèles à généraliser des relations pourtant élémentaires. ChatGPT-4 répond correctement à des questions comme « Qui est la mère de Tom Cruise ? [Réponse : Mary Lee Pfeiffer] » dans 79 % des cas, mais seulement dans 33 % des cas pour la formulation inverse « Qui est le fils de Mary Lee Pfeiffer ? »[35].

### Une société sous influence : qui pilote les algorithmes ?

L'entrée de l'I.A. dans notre quotidien s'accompagne d'une prise de conscience des risques liés à son utilisation. Une I.A. dont on perd le contrôle peut représenter une menace sur les individus et la société[36]. En fait, la question n'est pas tant de savoir si les êtres humains vont garder le contrôle sur l'I.A. mais plutôt s'ils vont garder le contrôle sur celles et ceux qui contrôlent l'I.A. Si les versions actuelles de l'I.A. permettent d'exploiter les biais cognitifs, de créer des faux comptes, de faire émerger de faux mouvements de soutien et de mettre en avant des contenus toxiques, ce n'est pas parce que l'I.A. a pris telle ou telle décision, mais parce que l'être humain derrière l'algorithme a rendu cela possible.

Aujourd'hui, des exemples récents nous amènent à réfléchir sérieusement à l'I.A. que nous voulons demain. Elon Musk contrôle le réseau *X* et a supprimé les équipes de modération. *X* est utilisé comme un outil de marketing pour promouvoir les entreprises de Musk, comme *Tesla* ou *SpaceX*, et de propagande pour diffuser des fausses informations au sein de l'opinion publique sur des sujets cruciaux comme la politique, l'innovation, les questions sociétales ou la crise climatique[37]. De la même façon, *TikTok* et *Facebook* (Meta) sont utilisés pour manipuler les décisions politiques, les choix d'achat et de consommation, et le comportement des personnes qui l'utilisent[38]. La question du contrôle de l'I.A. est donc la question du contrôle exercé par certains individus – qui détiennent un pouvoir considérable qui échappe à tout cadre réglementaire – sur d'autres individus.

## Construire l'I.A. de demain

Le contrôle de l'I.A. est une question trop sérieuse pour être laissée dans la main de quelques personnes dont les intérêts personnels balayent sans remords les intérêts collectifs. Les travaux de David Chavalarias (2023) mettent en garde contre les dangers potentiels de l'I.A. tout en appelant à une réflexion approfondie sur la manière dont ces technologies peuvent être conçues et utilisées de manière éthique et responsable. La régulation des plateformes numériques devient un enjeu fondamental pour préserver les fondements démocratiques et les droits fondamentaux[39]. La régulation ne se limite pas à imposer des règles techniques, mais doit incarner des choix éthiques, juridiques et politiques[40].

---

[33] https://www.nytimes.com/2025/05/29/well/maha-report-citations.html
[34] Shumailov *et al.*, 2024.
[35] Berglund *et al.*, 2023.
[36] Russell, 2019.
[37] Chavalarias, 2023.
[38] Coeckelbergh, 2023.

### Enjeu technique

Définir des règles techniques pour les algorithmes suppose la transparence des plateformes et de pouvoir faire des modifications et des corrections si cela s'avère nécessaire.

### Enjeu éthique

La communauté *fair-ML* qui œuvre pour des systèmes d'I.A. équitables et éthiques, suggère d'inclure les questions sociétales dès l'étape de conception technique[41]. Un consensus a émergé de la Communauté européenne, l'OCDE et l'UNESCO pour une I.A. éthique qui s'articule autour de 5 principes[42].

➢ Transparence et explicabilité : transparence sur les critères utilisés pour déterminer quels contenus sont promus ou supprimés et sur le fonctionnement des algorithmes.
➢ Justice et équité : l'I.A. doit prévenir la discrimination en évitant de perpétuer ou d'amplifier les biais existants, tels que les préjugés raciaux, de genre, socio-économiques ou culturels.
➢ Non-malfaisance : l'I.A. doit contribuer au bien-être des individus et de la société dans son ensemble, en évitant de promouvoir des contenus toxiques et nuisibles comme la haine, la violence, et la désinformation.
➢ Responsabilité et traçabilité : l'I.A. doit être conçue pour minimiser les erreurs, et des mécanismes doivent être en place pour les corriger. Il faut pouvoir identifier qui est responsable (concepteur, utilisateur, entreprise) en cas de problème. Les individus doivent avoir des moyens de contester les décisions d'une I.A., comme la suppression de contenus ou de comptes, et obtenir des réponses claires et rapides.
➢ Respect des droits fondamentaux : l'I.A. ne doit pas porter atteinte aux libertés individuelles ni aux libertés collectives (respect de la vie privée, protection des données personnelles, liberté d'expression, respect de l'environnement, etc.).

### Enjeu juridique

La régulation des plateformes numériques nécessite d'établir des normes et un cadre juridique pour le développement et l'usage de l'I.A., de définir les responsabilités et de mettre en place un système de sanctions.

### Enjeu politique

Parvenir à un consensus sur la régulation de l'I.A. nécessite une volonté politique nationale et internationale, comme celle qui a permis la mise en place du Règlement général sur la protection des données (RGPD) et de l'*A.I. Act*\* au niveau européen. Au-delà de la régulation, une volonté politique est également nécessaire pour agir sur l'éducation. Près de 70 % des 18-24 ans utilisent l'I.A. générative[43] et plus de 90 % des 16-24 ans utilisent au moins un réseau social[44]. Face aux risques évidents d'usage abusif, de désinformation et d'exploitation des biais cognitifs par ces plateformes, la question d'une politique éducative à la hauteur de la transformation sociétale qui a lieu est cruciale. Le système éducatif doit accompagner cette révolution numérique, pour enseigner au plus grand nombre à rester critique face aux images, vidéos et propos, et à différencier les sources sûres des fausses informations et récits de propagande 2.0[45].

---

[39] Awad *et al.*, 2018.
[40] Andler, 2023.
[41] Binns, 2018.
[42] Jobin *et al.*, 2019.
[43] Baromètre 2024, *Les Français et les I.A. génératives,* Vague 2 Ifop pour Talan, avril 2024.
[44] Eurostat (extraction du 16 janvier 2025), EU-TIC.
[45] Doshi *et al.*, 2024 ; Sun *et al.*, 2024.

Dans certains pays, comme la Finlande, l'éducation aux médias et aux technologies numériques a été intégrée dans le programme scolaire, dès l'école primaire. Les élèves apprennent à repérer les trolls, à interroger les images et à douter de ce qui est en ligne. Face aux risques de désinformation, l'éducation est un bouclier démocratique. C'est un choix politique que de soutenir l'éducation, un choix de société. À 6 ans, les enfants finlandais sont capables de repérer sur une photo, une incongruence. D'ailleurs, selon le *Media Literacy Index*[46] compilé par l'Open Society Institute-Sofia, la Finlande est classée comme le pays le plus résilient d'Europe face à la désinformation. Cet index compile différents indices sur la qualité de l'éducation, la liberté des médias, des indices de confiance interpersonnelle et d'engagement citoyen. En 2023, sur 41 pays européens, la France était au 17e rang.

# Conclusion

La place des réseaux sociaux et des I.A. génératives dans nos vies est croissante. Pour que cette révolution numérique prenne en compte le bien commun et les droits fondamentaux individuels, l'avènement de l'I.A. doit être accompagné de deux volets complémentaires d'actions : régulation et éducation.

Des actions de régulation sont essentielles face aux défis posés par la désinformation et le contrôle par une poignée d'individus – pour qui les droits fondamentaux sont parfois accessoires – de la quasi-totalité des plateformes numériques. Réguler l'I.A. est nécessaire pour protéger les droits humains, au niveau individuel et au niveau collectif. Le rôle des plateformes numériques dans la fabrication de vérités parallèles ne doit pas être sous-estimé[47]. Une idée naïve voudrait que, dans une démocratie, la vérité va résister à toutes les attaques qu'elle peut subir. Les données sur la multiplication des fausses informations et leur propagation montrent que cette idée est fausse. Il est aujourd'hui crucial de préserver l'autonomie et la liberté des individus face à l'influence croissante des algorithmes.

Des actions éducatives sont également essentielles. L'éducation constitue un levier fondamental : elle permet de renforcer la littératie numérique et de préparer les individus à faire des choix éclairés et critiques face aux technologies. Bien que les approches éducatives demandent du temps pour porter leurs fruits, elles sont indispensables et doivent s'articuler avec les dispositifs de régulation. Ensemble, éducation et régulation formeront une réponse cohérente et durable aux enjeux éthiques et sociétaux soulevés par l'I.A.

---

[46] « Bye, Bye Birdie: The Challenges of Disinformation », *The Media Literacy Index 2023*, https://osis.bg/wp-content/uploads/2023/06/MLI-report-in-English-22.06.pdf.
[47] Xun, J., 2024, « Hypnocratie : Trump, Musk et la fabrique du réel », *Philosophie Magazine*.

## Bibliographie

Ajwani, R., Javaji, S. R., Rudzicz, F., & Zhu, Z. (2024) *LLM-generated black-box explanations can be adversarially helpful,* arXiv:2405.06800v3

Andler, D. (2023), *Intelligence artificielle, intelligence humaine: la double énigme,* Paris, Gallimard

Awad, E., Dsouza, S., Kim, R., Schulz, J., Henrich, J., Shariff, A.,... & Rahwan, I. (2018), « *The moral machine experiment* ». *Nature*, 563(7729), p. 59-64

Berg, T., Burg, V., Gombović, A., & Puri, M. (2020),« *On the Rise of Fintechs: Credit Scoring Using Digital Footprints* », *Rev. Financ. Stud*, 33(7), p. 2845-2897

Berglund, L., Tong, M., Kaufmann, M., Balesni, M., Stickland, A. C., Korbak, T., & Evans, O. (2023) *The Reversal Curse: LLMs trained on" A is B" fail to learn" B is A",* arXiv:2309.12288

Bigman, Y. E., & Gray, K. (2018), « *People are averse to machines making moral decisions* », *Cognition*, 181, p. 21-34

Binns, R. (2018). « *Fairness in Machine Learning: Lessons from Political Philosophy* », *Proc FAT*, 81, p. 149-159

Birsch, D., & Fielder, J. H. (1994), *The Ford Pinto case: A study in applied ethics, business, and technology,* State University of New York Press

Bogacz, R., Brown, E., Moehlis, J., Holmes, P., & Cohen, J. D. (2006), « *The physics of optimal decision making: A formal analysis of models of performance in two-alternative forced-choice tasks* », *Psychol. Rev.*, 113(4), p. 700-765

Bruers, S., & Braeckman, J. (2014), « *A review and systematization of the trolley problem* ». *Philosophia*, 42, p. 251-269

Caliskan, A., Bryson, J. J., & Narayanan, A. (2017), « *Semantics derived automatically from language corpora contain human-like biases* », *Science*, 356(6334), p. 183-186

Chavalarias, D. (2023), *Toxic data,* Flammarion

Cheong, I., Xia, K., Feng, K. K., Chen, Q. Z., & Zhang, A. X. (2024), « *I am not a lawyer, but...: engaging legal experts towards responsible LLM policies for legal advice* » *Proc FAT*, p. 2454-2469

Coeckelbergh, M. (2023), « *Democracy, epistemic agency, and AI: political epistemology in times of artificial intelligence* », *AI and Ethics*, 3(4), p. 1341-1350

Doshi, J., Novacic, I., Fletcher, C., Borges, M., Zhong, E., Marino, M. C.,... & Xia, M. (2024), *Sleeper Social Bots: a new generation of AI disinformation bots are already a political threat,* arXiv:2408.12603

Emanuel, E. J., Persad, G., Upshur, R., Thome, B., Parker, M., Glickman, A.,... & Phillips, J. P. (2020), « *Fair allocation of scarce medical resources in the time of Covid-19* « , *NEJM*, 382(21), p. 2049-2055.

Eriksson, A., & Stanton, N. A. (2017), « *Takeover time in highly automated vehicles: noncritical transitions to and from manual control* », *Hum. factors*, 59(4), p. 689-705

Greene, J. D., Nystrom, L. E., Engell, A. D., Darley, J. M., & Cohen, J. D. (2004), « *The neural bases of cognitive conflict and control in moral judgment* », *Neuron*, 44(2), p. 389-400

Hassija, V., Chamola, V., Mahapatra, A. *et al.* (2024), « *Interpreting Black-Box Models: A Review on Explainable Artificial Intelligence* », *Cogn. Comput.* 16, p. 45–74

Haupt, C. E., & Marks, M. (2023), « *AI-generated medical advice – GPT and beyond* », *JAMA*, 329(16), p. 1349-1350

Jobin, A., Ienca, M., & Vayena, E. (2019), « *The global landscape of AI ethics guidelines* », *Nat. Mach. Intell.*, 1, p. 389–399

Kahneman, D., Slovic, P., & Tversky, A. (Eds.). (1982), *Judgment under uncertainty: Heuristics and biases,* Cambridge University Press

Koenigs, M., Young, L., Adolphs, R., Tranel, D., Cushman, F., Hauser, M., & Damasio, A. (2007), « *Damage to the prefrontal cortex increases utilitarian moral judgements* », *Nature*, 446(7138), p. 908-911

Lamb, W. F., Mattioli, G., Levi, S., Roberts, J. T., Capstick, S., Creutzig, F.,... & Steinberger, J. K. (2020), « *Discourses of climate delay* », *Glob. Sustain.*, 3, e17

Lang, Benjamin & Blumenthal-Barby, Jennifer & Nyholm, Sven. (2023), « *Responsibility Gaps and Black Box Healthcare AI: Shared Responsibilization as a Solution* », *Digit. Soc.*, 2. p. 1-18

Liu, F., Liu, Y., Shi, L., Huang, H., Wang, R., Yang, Z., ... & Ma, Y. (2024). *Exploring and evaluating hallucinations in llm-powered code generation,* arXiv:2404.00971

Marrapese, A., Suleiman, B., Ullah, I., & Kim, J. (2024), *A Novel Nuanced Conversation Evaluation Framework for Large Language Models in Mental Health,* arXiv:2403.09705

Moss-Racusin, C. A., Dovidio, J. F., Brescoll, V. L., Graham, M. J., & Handelsman, J. (2012), « *Science faculty's subtle gender biases favor male students* », *PNAS*, 109(41), p. 16474-16479

Palminteri, S., Garcia, B., & Qian, C. (2025), *How Objective Source and Subjective Belief Shape the Detectability and Acceptability of LLMs' Moral Judgments*, osf :10.31234/osf.io/ct6rx_v1

Pariser, E. (2011), *The Filter Bubble: How the New Personalized Web Is Changing What We Read and How We Think*, Penguin Press

Pemberton, S. A. (2016), *Harmful societies: Understanding social harm*, Policy Press

Ratcliff, R., & McKoon, G. (2008), « *The diffusion decision model: Theory and data for two-choice decision tasks* », *Neural Comput.*, 20(4), p. 873-922

Russell, S. (2019), *Human compatible: AI and the problem of control*, Penguin Press

Sejnowski, T. J. (2024), *ChatGPT and the Future of AI: The Deep Language Revolution*, MIT Press

Shumailov, I., Shumaylov, Z., Zhao, Y. *et al.* (2024), « *AI models collapse when trained on recursively generated data* », *Nature* 631, p. 755–759

Soltaninejad, K. (2020), « *Methanol mass poisoning outbreak, a consequence of COVID-19 pandemic and misleading messages on social media* », *Int. Jal Occup. Environ. Med.*, 11(3), p. 148-150

Spampatti, T., Hahnel, U.J.J., Trutnevyte, E. *et al.* (2024), « *Psychological inoculation strategies to fight climate disinformation across 12 countries* », *Nat. Hum. Behav.,* 8, p. 380–398

Sun, Y., He, J., Cui, L., Lei, S., & Lu, C. T. (2024), *Exploring the deceptive power of llm-generated fake news: A study of real-world detection challenges*, arXiv:2403.18249

Thomson, J. J. (1986), *Rights, restitution, and risk: Essays, in moral theory*, Harvard University Press

Vosoughi, S., Roy, D., & Aral, S. (2018), « *The spread of true and false news online* », *Science*, 359(6380), p. 1146-1151